\input{aipcheck}

\documentclass[
    ,final            
  ]
  {aipproc}

\layoutstyle{6x9}

\begin{document}

\title{Coulomb distortion in the inelastic regime
\footnote{This work was supported by U.S. Department of Energy, Office of Nuclear Physics, 
under contract numbers DE-AC02-06CH11357 and DE-AC05-84ER40150 Modification No.~M175.}}


\author{P. Solvignon}{
  address={Physics Division, Argonne National Laboratory, Argonne, IL 60439}
}

\author{D. Gaskell}{
  address={Thomas Jefferson National Accelerator Facility, Newport News, VA 23606}
}

\author{J. Arrington}{
  address={Physics Division, Argonne National Laboratory, Argonne, IL 60439}
}

\begin{abstract}
The Coulomb distortion effects have been for a long time neglected in deep inelastic 
scattering for the good reason that the incident energies were very high. But for energies
in the range of earlier data from SLAC or at JLab, the Coulomb distortion could have 
the potential consequence of affecting the A-dependence of the EMC effect and of the 
longitudinal to transverse virtual photon absorption cross section ratio $R(x,Q^2)$.
\end{abstract}

\maketitle

%
\section{Introduction}
The incident (scattered) electron is accelerated (deccelerated) due to the Coulomb field 
created by the nearby protons through the exchange of soft photons. The Coulomb distortion 
manifests when the electomagnetic interaction is used to probe the nucleus and its effect 
increases with the number ($Z$) of protons inside the nucleus. The effect is expected to 
have the opposite sign with a positron beam. 

Due to these changes in the initial and final kinematics and the wavefunctions of the 
lepton, the cross sections cannot be expressed by the Plane Wave Born Approximation 
(PWBA); the Distorded Wave Born Approximation (DWBA) should be considered instead. 
However full calculation of DWBA is very extensive and difficult to implement. A more 
convenient approach is the Effective Momentum Approximation (EMA) where the incident
and scattered electron energies ($E$ and $E_p$, respectively) are shifted by an average 
Coulomb potential, $\bar{V}$,
seen by the electron. Moreover, a focusing on the incoming lepton plane wave needs to be 
applied. In Ref.~\cite{Aste:2005wc}, A.~Aste and collaborators have made a detailed 
comparison of the EMA to a full DWBA calculation for quasi-elastic scattering. In their 
approach, the spectral function is subject to the transformation 
$S_{tot}^{PWBA}(|\vec{q}|,\omega,\theta) \longrightarrow S_{tot}^{PWBA}(|\vec{q}_{eff}|,\omega,\theta)$, with $\omega$ and $\vec{q}$ the energy and momentum transferred to the nucleon and $\theta$ the 
scattering angle.
The Mott cross section undergoes the transformation $\sigma_{Mott} \longrightarrow 
\sigma_{Mott}^{eff}$ with:
\begin{eqnarray}
\sigma_{Mott}^{eff} = 4 \alpha^2  \cos^2(\theta/2) (E_p+\bar{V})^2 / Q^4_{eff}
\label{mott_eff}
\end{eqnarray}
with the effective squared four-momentum transfer $Q^2_{eff}$ equal to $4 (E+\bar{V})(E_p+\bar{V}) \sin^2(\theta/2)$, $\alpha$ the fine structure constant.
Finally, the incoming focusing factor $F^i_{foc}$ is equal to $(E + \bar{V})/E$.
The resulting cross section has the following form:
\begin{eqnarray}
\sigma_{tot}^{CC} = (F^i_{foc})^2 \cdot \sigma_{Mott}^{eff} \cdot S_{tot}^{PWBA}(
|\vec{q}_{eff}|,\omega,\theta)
\label{sig_cc2}
\end{eqnarray}
The electrostatic potential $V(r)$ inside the charged sphere of radius 
$R = 1.1 A^{1/3} + 0.86 A^{-1/3}$ defined as followed:
\begin{eqnarray}
V(r) = -\frac{3 \alpha (Z-1)}{2 R} + \frac{\alpha (Z-1)}{2 R} \left(\frac{r}
{R}\right)
\label{e2}
\end{eqnarray}
where the effect of the probed proton is neglected. Because most the nucleons of heavy 
nuclei are located in the nucleus peripherical region, taking the electrostatic potential 
at the center of the nucleus will be an overestimate of the Coulomb effect. When using an 
effective potential $\bar{V} \approx (0.75...0.80) V(0)$, the authors in~\cite{Aste:2005wc} 
showed the good agreement of EMA with the full DWBA calculation~\cite{Jin:1994zw}. 
The same effective potential value was extracted from positron and electron quasi-elastic 
scattering experiment~\cite{Gueye:1999mm}.
\section{Impact on the EMC effect}
\begin{figure}[t]
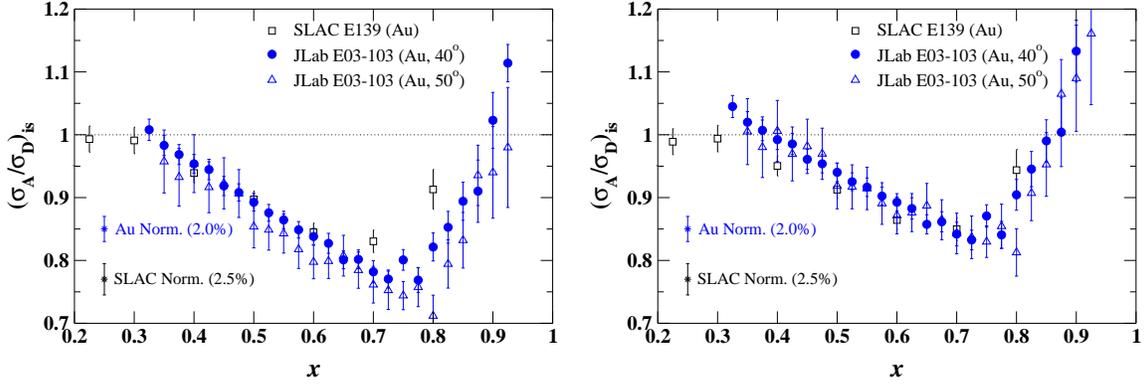

  \includegraphics[height=.23\textheight]{coulomb_au_nocc_bis.eps}
  \hspace{0.2cm}
  \includegraphics[height=.23\textheight]{coulomb_au_cc_bis.eps}
  \label{fig_au}
  \caption{The EMC effect in Gold before (left plot) and after (right plot)
           Coulomb corrections (corrections have been applied to all data sets). 
           The JLab E03-103 data are preliminary.}
\end{figure}
The EMC effect was first observed at CERN when physicists decided to use heavy 
nuclei in order to increase the luminosity in the study of nucleon substructure. 
They discovered that the quark distributions are different when the nucleon is inside 
the nuclear medium. Many other experiments confirmed this observation, and even with
more complete and precise data from SLAC E139~\cite{Gomez:1993ri}, the origin of this
reduction of the high momentum quark distribution in nuclei is not yet well understood.
It is widely accepted that nuclear binding and Fermi motion components must be included
in realistic theoretical calculations, although they alone cannot explain the 
EMC effect. More {\it exotic} ideas may need to be included, e.g. nuclear pions, 
multiquark clusters, dynamical rescaling, etc (see Ref.~\cite{Geesaman:1995yd} for 
a review).

New results from JLab experiment E03-103~\cite{Seely:2009gt} on heavy nuclei 
motivated us to have a more detailed look at the possible significant effect from
the Coulomb distortion. Unfortunately there have been no detailed calculations of the 
Coulomb distortion in the inelastic regime for beam energies on the order of several GeV 
({\it i.e. SLAC kinematics}). In particular there are no studies verifying that EMA is a good 
approximation in this channel. Nonetheless the EMA is a reasonable approach for making 
an estimate regarding the potential impact of Coulomb distortion in inelastic scattering.
Using the formalism defined in the previous section, the data from E03-103 were corrected 
as shown in Fig.~\ref{fig_au} for Gold. Upward shifts of about 10\% and 3\% are observed 
for the JLab E03-103 and SLAC E139 data respectively. The slight difference between the 
two E03-103 data sets seems to be resolved after the correction. This may indicate that 
the use of EMA is valid in the inelastic regime too. Note, though, that the apparent 
agreement between the JLab and SLAC data gets worse after corrections. This will be address 
in more details later. It can be clearly seen that neglecting the effect of the Coulomb 
field at the SLAC or JLab energies would imply an overestimate of the EMC effect in 
medium-weight to heavy nuclei.

From measurements of the EMC effect on a wide range of nuclei, one can extrapolate to an 
infinitely heavy nucleus, also called infinite nuclear matter. These results are key to 
our understanding of nucleon's in-medium modification as they can be compared to exact 
calculations~\cite{Smith:2003hu,Cloet:2006bq} and avoid the complexity arising from the 
nuclear structure of the finite nuclei. A global analysis of the world 
data~\cite{Solvignon:2009nm} has been performed which includes Coulomb correction. The 
data most sentisitive to this analysis are the SLAC E139 and JLab E03-103 data which were 
taken at relatively lower energies than CERN. Because the Coulomb distortion increases 
with $Z$ ($A$), their effects change the extrapolation from finite nuclei to infinite 
nuclear matter.
\begin{figure}[t]
  \includegraphics[height=.26\textheight]{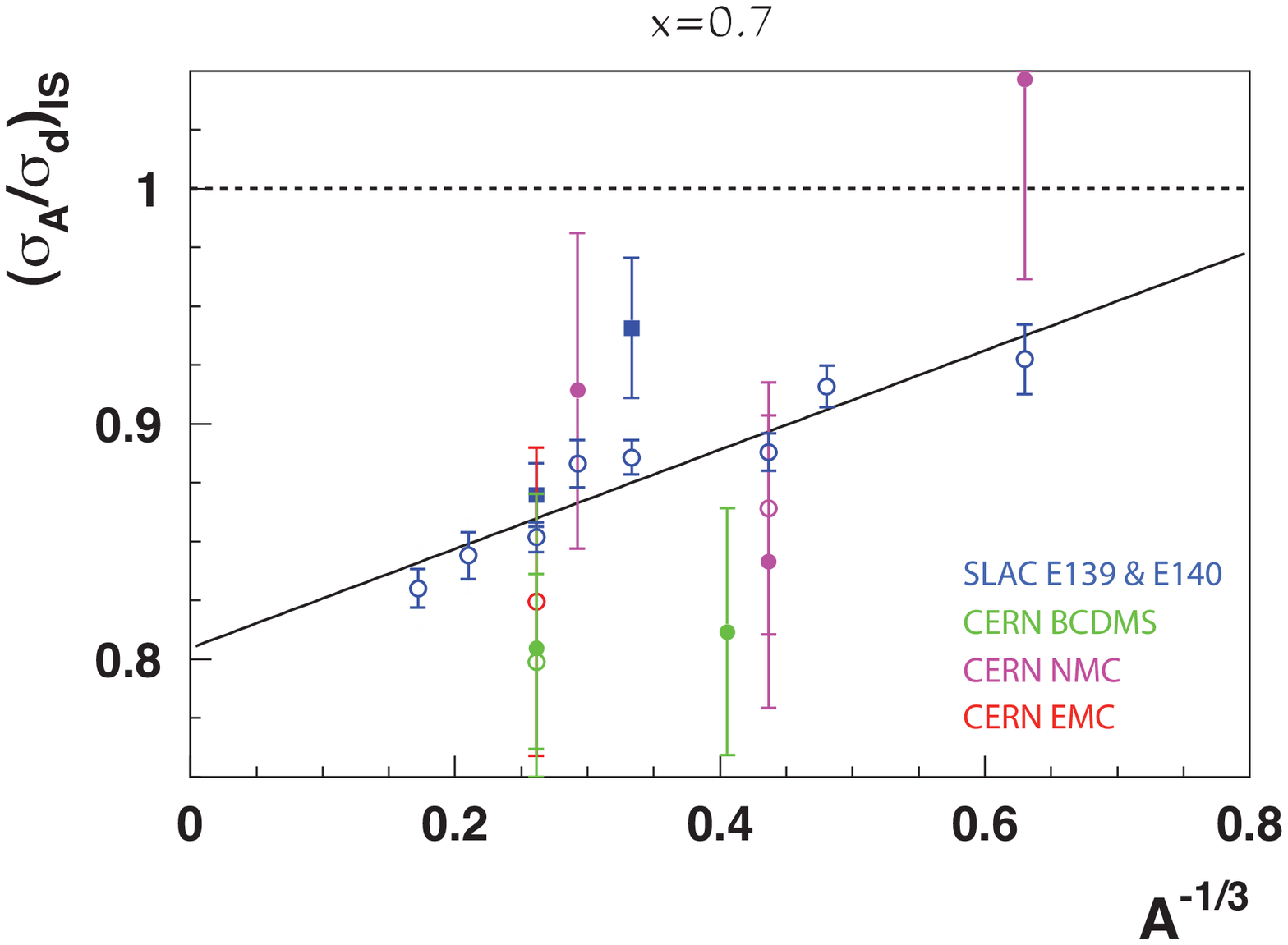}
  \includegraphics[height=.26\textheight]{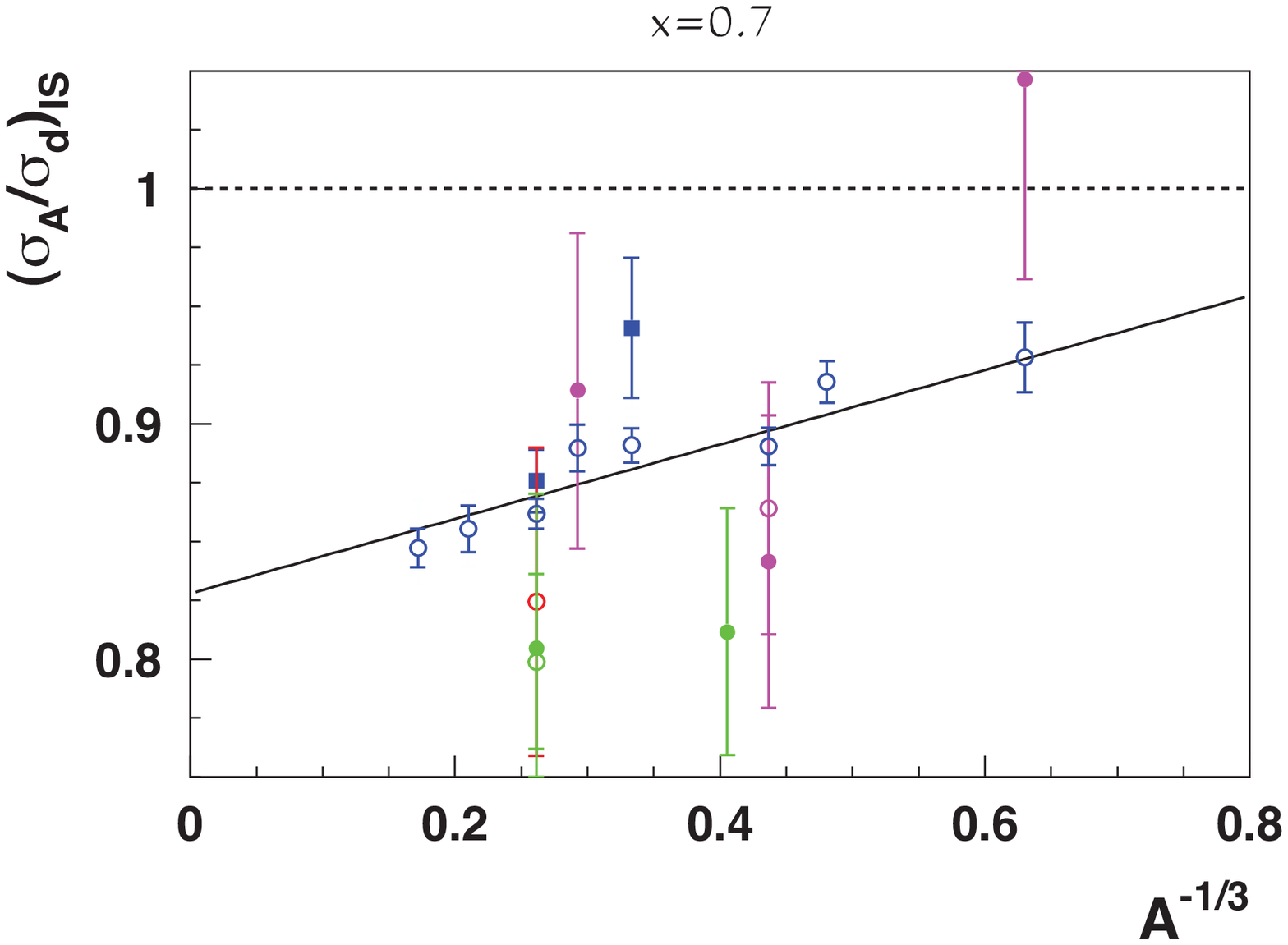}
  \label{fig_nm}
  \caption{Extrapolation to infinite nuclear matter before (left plot) and after 
           (right plot) Coulomb correction. The lines are fits of all the data plotted.}
\end{figure}

Following the same procedure as in Ref.~\cite{Sick:1992pw}, the size of the EMC effect 
in infinite nuclear matter was extracted at several $x$-values in order to map out its 
$x$-dependence. In Fig.~\ref{fig_nm}, the extrapolations from the original world data 
and from the Coulomb corrected ones are illustrated for $x = 0.7$. A 3\% difference is 
found between the two extrapolations. Including the preliminary Coulomb corrected data 
from E03-103 does not change the extrapolated value for nuclear matter. This gives us 
some confidence that the Coulomb distortion method used is appropriate as well in the 
deep inelastic scattering regime, given the large (-10\%) correction for Gold at 
$x \simeq 0.75$. On a related topic, other dependences of the EMC effect are being looked 
at, e.g. with more involved nuclear density calculations or using the nucleon overlap 
distributions~\cite{Wiringa_priv}. 
\section{The nuclear dependence of $R=\sigma_L/\sigma_T$ revisited}
The longitudinal to transverse virtual photon absorption cross section ratio $R$ contains 
important information on the nature of the nucleon's constituents~\cite{Geesaman:1995yd}. 
If the nucleon is composed mostly of spin-1/2 partons, $R$ ($R \sim 4 m^2_q/Q^2$) should be 
very small ($\simeq 0$) and decrease with $Q^2$. However, if spin-0 partons and/or 
correlations between partons have non-negligible contributions, $R$ will be larger and 
increase with $Q^2$. 

Measuring $R$ for nuclear targets is of equal importance as it could shed light on the 
underlying physics responsible for the EMC effect. A demonstration of the nuclear 
independence of $R$ would lead to the conclusion that confinement properties are the same 
in a free nucleon and in a bound nucleon. If a nuclear dependence would be observed, one 
could state that the nuclear medium is changing the contribution of the spin-0 partons and 
higher twist effects of the nucleon.

The most precise test of the A-dependence of $R$ at large $x$ was done in~\cite{Dasu:1993vk}.
The authors found no nuclear dependence of $R$ and therefore concluded that 
$\sigma_A/\sigma_D$ can be directly used to study the nuclear dependence of the EMC ratio, 
which is defined as $F_2^A/F_2^D$.
\begin{figure}[t]
  \includegraphics[height=.25\textheight]{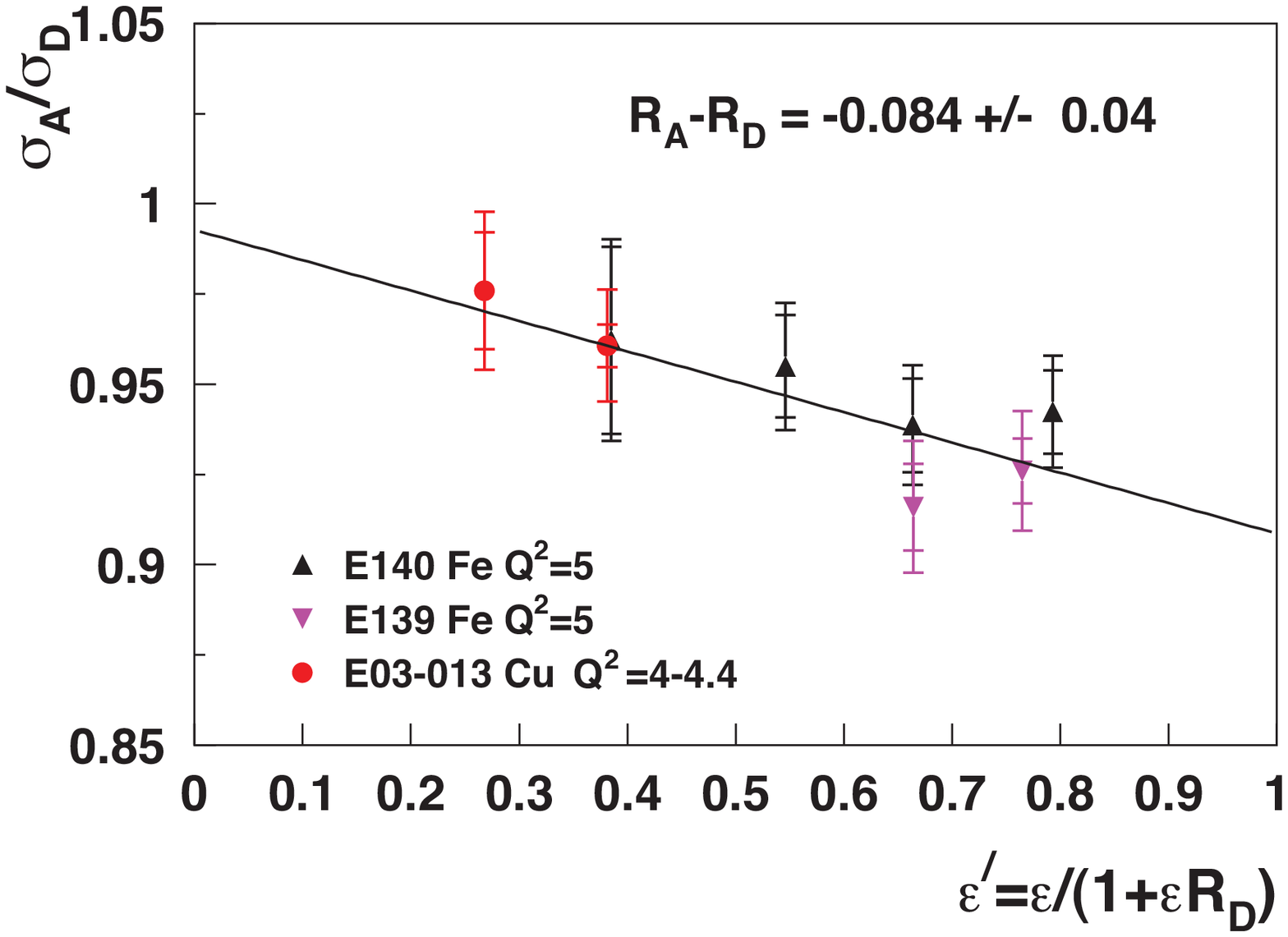}
  \hspace{0.5cm}
  \includegraphics[height=.28\textheight]{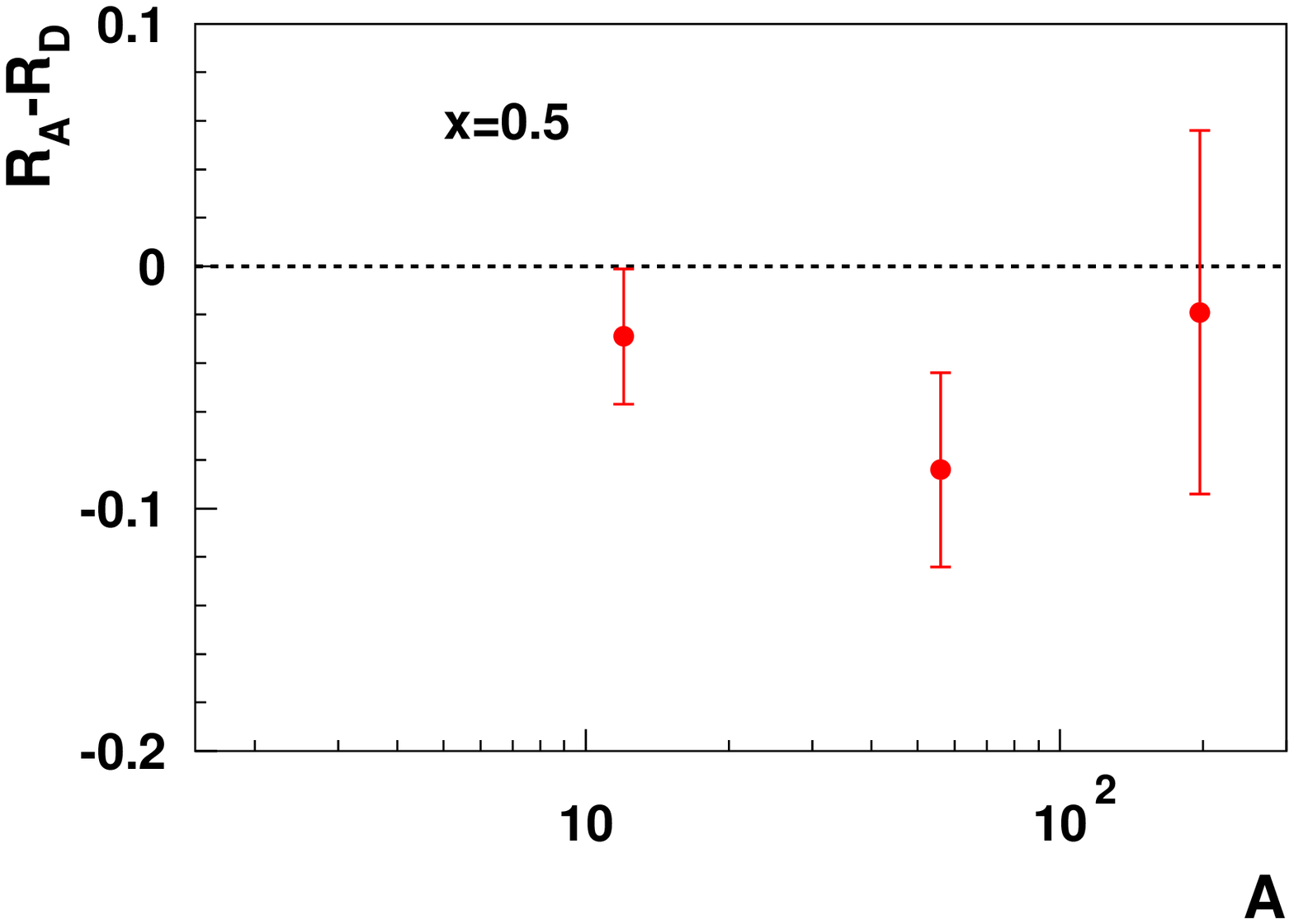}
  \label{fig_ra}
  \caption{Extraction of the nuclear dependence of $R$ for $^{56}$Fe-$^{63}$Cu at $x =0.5$ 
           (left plot). Results from the same extraction method for $^{12}$C, 
           $^{56}$Fe-$^{63}$Cu and $^{197}$Au (right plot). 
           The JLab E03-103 data are preliminary.}
\end{figure}
However in the analysis of Ref.~\cite{Dasu:1993vk}, the Coulomb distortion effects have been 
assumed to be small. A re-analysis of these data including Coulomb 
corrections~\cite{Gaskell:2009ra} shows a non-trivial change of slope in the extraction of 
$R_A - R_D$ (from -0.018 $\pm$ 0.055 to -0.051 $\pm$ 0.058), but being still consistent with
zero within the error bar.. This observation becomes even 
more obvious after including the data from~\cite{Gomez:1993ri} and the preliminary data on 
Copper from JLab E03-103 (see left plot in Fig.~\ref{fig_ra}), which gives us a larger lever 
arm in $\epsilon'= \epsilon/(1+\epsilon R_D)$. The value for $R_A - R_D$ at $x = 0.5$ is 
then found to be -0.084 $\pm$ 0.040. The use of variable $\epsilon'$ is convenient to extract 
the A-dependence of $R$ as, in the case of $R_A = R_D$, $\sigma_A/\sigma_D$ should be found 
constant when plotted versus $\epsilon'$. Before applying the corrections, $R_A - R_D$ from 
the same set of data was equal to -0.035 $\pm$ 0.040.
In fact, Coulomb corrections are strongly $\epsilon$-dependent and therefore corrupt the 
$\epsilon$-dependence in the extraction of $R$ in medium to heavy nuclei. 

Finally, note that this apparent nuclear dependence of $R$ explains the discrepancy between 
the E139 and JLab data after Coulomb corrections.
\section{Conclusion}
Coulomb distortion cannot any longer be considered as a low energy effect. It was 
shown in this paper that, for earlier SLAC experiments, the Coulomb corrections were 
small but not negligible. For JLab E03-103 data, the corrections can certainly not 
be disregarded. The agreement between the SLAC and JLab data is very good after this 
correction is applied and the extrapolation to infinite nuclear matter is shifted 
upward by about 3\% relative, which is a significant effect when one wants to compare
to more evolved theoretical calculations. 

It is also not clear whether the longitudinal to transverse virtual photon absorption 
cross section ratio $R$ is nuclear independent or not, and a more precise measurement 
is needed. More generally, all measurements using medium-weight to heavy nuclei are sensitive 
to the Coulomb distortion, and it is therefore crucial to perform measurements of the Coulomb
potential in different lepton scattering regimes, which requires the direct comparison of
inelastic scattering data using both electron and positron beams.
\section{Positron beam requirements}
A possible measurement of the Coulomb distortion in the inelastic region will require a 
positron beam intensity of at least 10 nA for a basic check and about 100nA for complete 
and detailed measurement in a reasonable beamtime of about a week.






\bibliographystyle{aipprocl} 

\bibliography{couldist_jpos09}

\end{document}